\documentclass[prl,floatfix,twocolumn,aps,showpacs,preprintnumbers,amsmath,amssymb]{revtex4}
\usepackage[latin1]{inputenc}

\usepackage{ulem} % For correction remarks

\usepackage{graphicx}% Include figure files
\usepackage{psfrag}
\usepackage{wasysym}
\graphicspath{{./Figures/}}

\newcommand{\be}{\begin{equation}}
\newcommand{\ee}{\end{equation}}
\newcommand{\bea}{\begin{eqnarray}}
\newcommand{\eea}{\end{eqnarray}}
\def\bea{\begin{eqnarray}}
\def\eea{\end{eqnarray}}
\def\ben{\begin{equation}}
\def\een{\end{equation}}
\def\benu{\begin{enumerate}}
\def\enu{\end{enumerate}}

\def\sss{\scriptscriptstyle\rm}

\def\1var{(\bx_1...\bx\N)}

\def\br{{\bf r}}

\def\bx{{\br t}}

\def\N{_{\sss N}}

\def\ee{_{\rm ee}}

\def\sph_int{ {\int d^3 r}}

\input rotate

\begin{document}

\title{Multifractality at the quantum Hall transition: Beyond
  the parabolic paradigm}

\author{F. Evers$^{1,2}$, A. Mildenberger$^3$, and A. D. Mirlin$^{1,2,*}$}
\affiliation{
\mbox{$^1$Institut f\"ur Nanotechnologie, Forschungszentrum Karlsruhe,
 76021 Karlsruhe, Germany}\\
\mbox{$^2$Institut f\"ur Theorie der kondensierten Materie,
 Universit\"at Karlsruhe, 76128 Karlsruhe, Germany}\\
\mbox{$^3$Fakult\"at f\"ur Physik,
 Universit\"at Karlsruhe, 76128 Karlsruhe, Germany}
}

\date{\today}

\begin{abstract}
We present an ultra-high-precision numerical 
study of the spectrum of multifractal
exponents $\Delta_q$ characterizing anomalous scaling of wave function 
moments $\langle|\psi|^{2q}\rangle$ at the quantum Hall transition. The 
result reads $\Delta_q = 2q(1-q)[b_0 + b_1(q-1/2)^2 + \ldots]$,
with  $b_0 = 0.1291\pm 0.0002$ and $b_1 = 0.0029\pm 0.0003$. 
The central finding is that the spectrum is not exactly parabolic, $b_1\ne 0$.
This rules out a class of theories of Wess-Zumino-Witten type proposed
recently as possible conformal field theories of the quantum Hall critical
point.

\end{abstract}

%%\pacs{72.15.Rn, 05.45.Df}

\maketitle

\paragraph*{Introduction} The quantum Hall effect is a famous
macroscopic quantum phenomenon \cite{vonKlitzing80,prange87} whose discovery
gave rise to one of the most active research areas in condensed matter physics 
of last three decades. 
The plateaus with quantized values of the Hall conductivity are separated by
quantum Hall transitions, which represent a celebrated example of a quantum
critical point in a disordered electronic system (for a recent review, see 
Ref.~\onlinecite{evers07}).  Identification of the critical field theory of 
the integer quantum Hall transition remains a
major unsolved problem of condensed matter physics. 

One of the key characteristics of the quantum Hall transition point is the
multifractality spectrum governing fluctuations of amplitudes of critical 
wave functions. Specifically, the moments of
wavefunctions scale with system size, $L$, 
with a set of anomalous exponents $\Delta_q$, 
\begin{equation}
\label{e1}
    \langle |\psi({\bf r})|^{2q} \rangle /  
\langle |\psi({\bf r})|^{2} \rangle^q
 \sim L^{-\Delta_q}\ .
\end{equation}
(The angular brackets denote the ensemble averaging.)
Equivalently, one often characterizes multifractality by 
a closely related set of exponents, 
$\tau_q \equiv d(q-1)+\Delta_q$, or by its Legendre transform $f(\alpha)$
(``singularity spectrum''), $\alpha_q = \tau'_q$, 
$f(\alpha_q) = q\alpha_q - \tau_q$.  
Here $d$ is the system dimensionality (while $d=2$ for the present case of
the quantum Hall transition, we find it useful to keep it as $d$ in
formulas below), and the prime denotes the $q$-derivative.  

Zirnbauer\cite{zirnbauer99} and Tsvelik\cite{bhaseen00,tsvelik07} conjectured
that the conformal theory of the quantum Hall critical point is of the
Wess-Zumino-Witten type. These proposals imply\cite{note0} 
that the spectrum $\Delta_q$ is
parabolic, i.e., that $\gamma_q$ defined according to  
\begin{equation}
\label{e2}
    \Delta_q = \gamma_q q (1-q)\:. 
\end{equation}
is in fact $q$-independent\cite{note1}, $\gamma_q=\gamma$. 
An accurate numerical analysis
of the multifractal spectrum plays therefore a crucial role for identification
of the critical theory. 

A high-accuracy evaluation of the multifractality spectrum was carried out in
our earlier work \cite{evers01}. For this purpose, we modelled systems of a
much larger size than in preceding works and performed averaging over
a large ensemble of wave functions, as well as a thorough analysis of
finite-size effects. It was found that the spectrum is close-to-parabolic,
$\Delta_q \simeq \gamma q(1-q)$ with $\gamma = 0.262 \pm 0.003$, thus showing
that, if deviations from parabolicity are present, they are rather small
(of the order of $1\%$).  While the data of Ref.~\onlinecite{evers01} were showing 
some indications for such deviations,  
they were smaller than the numerical uncertainties. 
The latter originate from statistical noise (limited size of the data
set) and from finite size effects affecting the scaling relation (\ref{e1})
which is used to extract $\Delta_q$.

The goal of the present Letter is to determine the $\Delta_q$ spectrum with an
ultra-high precision and to give an ultimate answer on the question ``Is the
multifractality spectrum of the quantum Hall transition strictly parabolic?''
For this purpose, we improve upon the earlier numerical analysis 
in two different ways. First, we utilize a statistical ensemble that
contains approximately ten times more samples than the one used
before\cite{evers01}. 
Second, we employ a recently discovered \cite{mirlin06} 
``reciprocity relation'', 
\begin{equation}
\label{e3}
\Delta_q = \Delta_{1-q},
\end{equation}
for a better control of finite-size corrections.  

Relation (\ref{e3}) implies a symmetry of the $\Delta_q$ spectrum 
around the point $q{=}1/2$. Consequently, 
an expansion of $\gamma_q$ about this point has a form
\begin{equation}
    \gamma_q/d = b_0 + b_1(q-1/2)^2 + b_2(q-1/2)^4\ldots.
\label{e3a}
\end{equation}
To verify or to exclude parabolicity of $\Delta_q$, the prefactor $b_1$ of the
quadratic  term (and possibly those of higher order terms) 
in Eq. (\ref{e3a}) should be determined numerically. 
This is the purpose of the present work. We will provide numerical evidence
that the corrections to parabolicity do not vanish. Specifically, we 
obtain $b_0{=}0.1291\pm 0.0002$ and $b_1{=}0.0029\pm 0.0003$, the non-zero
$b_1$ implying that the parabolcity is not exact. The corresponding value of
$\alpha_0$ [position of the apex of the singularity spectrum $f(\alpha)$] is 
$\alpha_0 = d + \gamma_0 = 2.2596 \pm 0.0004$.

\paragraph*{Method:} In order to find the critical eigenstates, 
we employ the same numerical strategy
that has been developed before\cite{evers01}. We determine the
lattice time  evolution operator $U$ for the Chalker-Coddington 
network model\cite{chalker87,klesse99} with periodic boundary
conditions and $N{=}2L^d$ nodes. Eight eigenstates with eigenvalues 
closest to unity are found
with a standard sparse matrix package\cite{num1,num2,num3} from
exact diagonalization of $U$. We study systems with 
$L{=}16, 32, 64,\ldots, 1024$ with $\sim 10^6$ samples for the 
smallest sizes and $\sim 10^4$ for the largest ones. 
% L	bulk		surface
% 16	2048000		2048000
% 32	1024000		1024000
% 64	1024000		1024000
% 128	555308		461397
% 192	212519		205222
% 256	120330		134974
% 384	52652		59045
% 512	29462		3393 (here I've lost a node at some point..)
% 768	12832		12515
% 1024	7909		7782
The statistical analysis proceeds via
calculating the average inverse participation ratios,
\begin{equation}
    P_q = \left\langle \int |\psi^2|^q \right\rangle 
\label{e5}
\end{equation}
which obey the scaling law
\begin{equation}
   P_q = c_q(N) \ N^{-(q-1)-\Delta_q/d}.
\label{e6}
\end{equation}
The coefficients $c_q$ become independent of $N$ in the limit 
$N\to\infty$. 

As an alternative approach, we consider the ratio 
\begin{equation}
\label{e7}
L_q=\frac{\langle \int |\psi^{2}|^q\ \ln |\psi|^2 \rangle}{\langle \int
|\psi^2|^q \rangle} \equiv (\ln P_q)'
= - {\alpha_q \over d} \ln N + (\ln c_q)'\,,
\end{equation}
whose scaling yields the exponent $\alpha_q = \Delta_q^\prime + d$. 
In this way, the exonent $\alpha_q$ is studied
directly, i.~e. without invoking a numerical
differentiation which can significantly increase the error bars.
In analogy with Eq.~(\ref{e3a}), we can
expand  $\Delta_q^\prime$ around $q=1/2$, 
\begin{equation}
\Delta_q^\prime = (1-2q)\ \tilde\gamma_q, \qquad 
\tilde\gamma_q/d =a_0 + a_1(q-{1\over 2})^2 + \ldots.
\label{e9}
\end{equation}
The coefficients of both expansions are related via 
$a_0{=}b_0{-}b_1/4$, $a_1{=}2b_1{-}b_2/2$, \ldots . 

The averages entering Eqs.~(\ref{e5}) and (\ref{e7})
are readily obtained numerically. It is beneficial to perform the scaling 
analysis of the ratio (\ref{e7}) in addition to that of Eq.~(\ref{e5}) 
for several reasons. First, 
the curvature of $\Delta_q$ is more clearly seen in the
$q$-derivative, $\Delta^\prime_q$. 
Second, the finite-size corrections are different in the cases
of Eqs.~(\ref{e5}) and (\ref{e7}), so that 
an agreement between the obtained exponents provides 
an additional confirmation of the validity of the $N\to\infty$ extrapolation
procedure. Also, the relation $a_1{=}2b_1{-}b_2/2$
allows one to extract the coefficient  $b_2$ of the quartic term in
Eq.~(\ref{e3a}) out of parabolic fits for $\gamma_q$ and $\tilde{\gamma}_q$.

%%%%%%%%%%%%%%%%%%%%%%%%%%%%%%%%%%%%%%%%%%%%%%%%%%%%%%%%%%%%%%%%%%%%
\begin{figure}[htb]
\begin{center}
\includegraphics[width=1.0\linewidth]{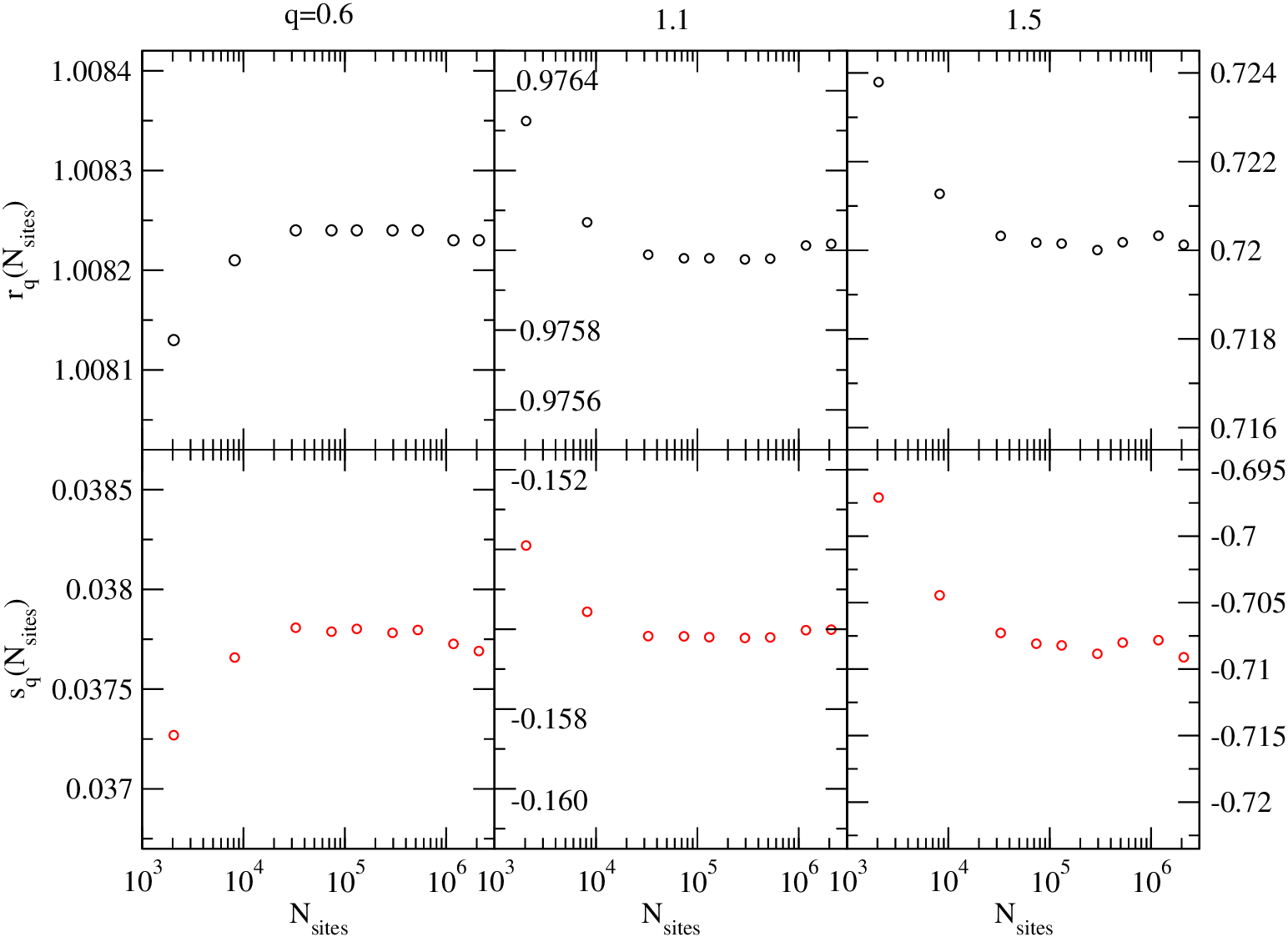}
\end{center}
\caption{Upper row: ratio $r_q$ of inverse participation numbers $P_q$ and
  $P_{1-q}$ [Eq.~(\ref{e10})] at $q{=}0.6,1,1.5$ (from left to
  right). The flat asymptotics indicates validity of the reciprocity
  relation, $\Delta_q{=}\Delta_{1-q}$. Small deviations from the constant
  behavior at largest system sizes are due to residual 
statistical noise. Lower row: analogous plots
for the logarithmic derivative $s_q = (\ln r_q)'$ defined in Eq.~(\ref{e11}).  
%Again, the flat asymptotics
%  exhibits the asymptotic reciprocity also for typical averages 
%  (\ref{e7}) inside the numerical window.
}
\label{f1}
\end{figure}
%%%%%%%%%%%%%%%%%%%%%%%%%%%%%%%%%%%%%%%%%%%%%%%%%%%%%%%%%%%%%%%%%

\paragraph*{Numerical results:} 
We begin the analysis of our numerical results by verifying the 
reciprocity relation (\ref{e3}). To this end, we consider the ratio
\begin{equation}
\label{e10}
    r_q{=}N^{2q-1}\frac{P_q}{P_{1-q}} = N^{(\Delta_{1-q}-\Delta_q)/d}
{ c_q(N) \over c_{1-q}(N)}\:.
%%    r_q{=}N^{1-2q}P_q/P_{1-q}
\end{equation}
The reciprocity relation (\ref{e3}) implies that the
leading powers should cancel, 
so that $r_q$ exhibits only subleading corrections in
$1/N$.  The log-linear plot, Fig.~\ref{f1}, upper row,
shows that $r_q$
saturates in the large $N$-limit with a very well defined asymptotic
value. Thus, we confirm reciprocity for the exponent spectrum of the
integer quantum Hall effect, as expected. \cite{mirlin06} 
Since the exponent relation must hold only in the asymptotic
regime, we can draw another conclusion which is important for the
subsequent analysis: the observed saturation of $r_q$ provides evidence that 
our numerically accessible sample sizes are indeed large enough in order to
be able to study the true asymptotics. 

Similar to $r_q$, we consider the logarithmic derivative
\begin{equation}
\label{e11}
s_q \equiv (\ln r_q)' = 
{2d-\alpha_q - \alpha_{1-q} \over d} \ln N + (\ln c_q c_{1-q})', 
\end{equation}
which also saturates well inside the numerical window, see
Fig.~\ref{f1}, lower row. Thus, the true asymptotics of $\alpha_q$ 
may be studied by means of Eq. (\ref{e7}) with available system sizes, too.
 
%\footnote{We mention, that the combined
%information contained in $r_q$ and $s_q$ provides a means to determine the
%coefficients $c_q$ with great accuracy. However, this is out of the
%scope of our present work.} 

%%%%%%%%%%%%%%%%%%%%%%%%%%%%%%%%%%%%%%%%%%%%%%%%%%%%%%%%%%%%%%%%%%%
\begin{figure}[htb]
\begin{center}
\includegraphics[width=1.0\linewidth]{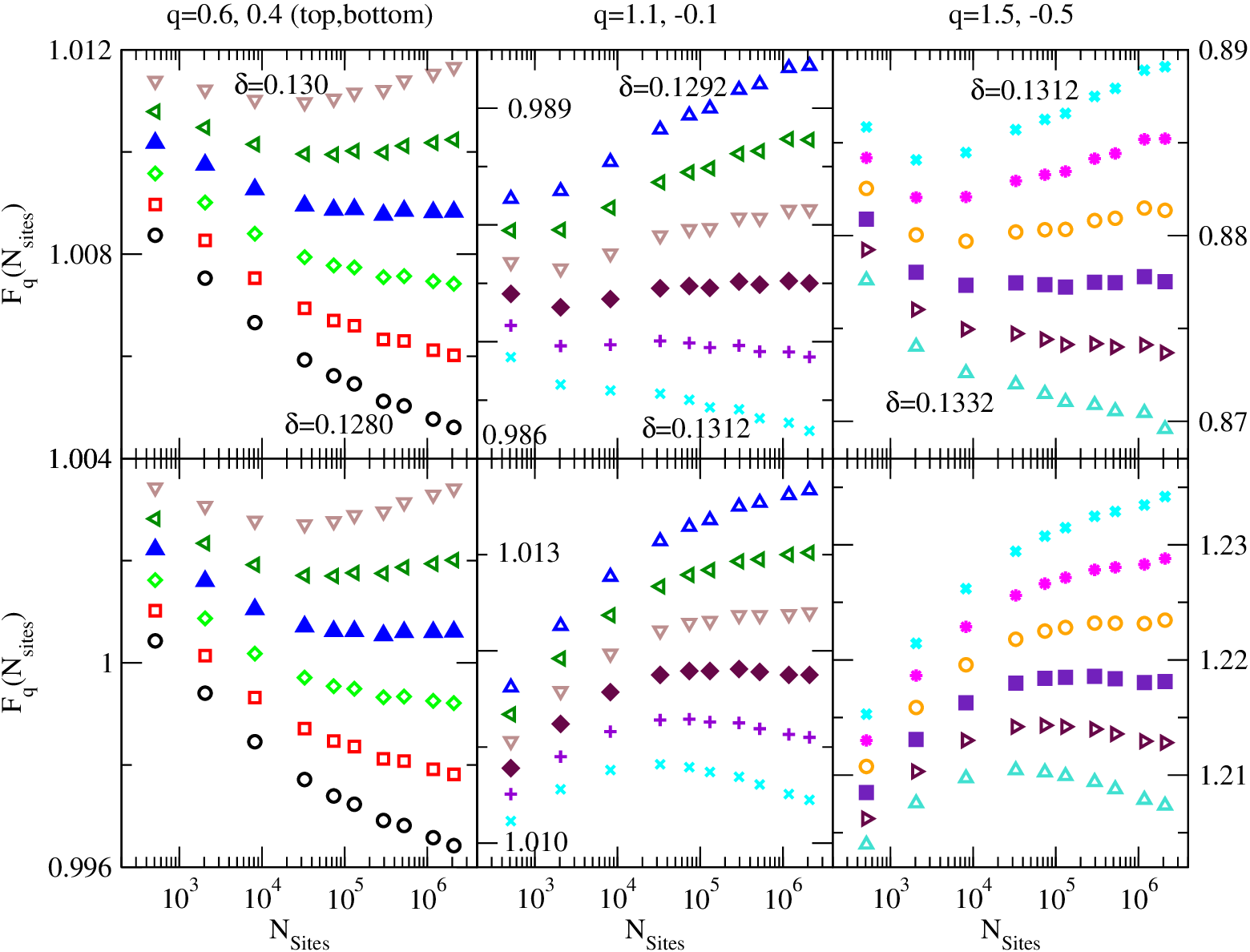}
\end{center}
\caption{Family of curves $F_q(N)= P_q N^{q-1 - \delta(q-1)q}$ 
for $q{=}0.6,1.1,1.5$ (top; left,center,right) and 
$q{=}0.4,-0.1,-0.5$ (bottom; left,center,right).
Each data set is labelled by a parameter $\delta$, 
which increases from a minimum to a maxium value
(given in the upper plots) in steps of 0.0004.
The value of $\delta$ for which $F_q(N)$ is flat determines the anomalous
exponent,  $\Delta_q{=}d \delta q(1-q)$.  Such saturating data sets
are marked with filled symbols (left: $\triangle$; 
center: $\Diamond$; right: $\Box$); typical error in
the corresponding value of $\delta$ does not exceed $0.001$. 
The change in symbols (i.e. in $\delta$) for saturating data sets 
illustrates a $q$-dependence of $\gamma_q$ and thus gives direct, unprocessed 
evidence of nonvanishing quartic terms in $\Delta_q$.}
\label{f2}
\end{figure}
%%%%%%%%%%%%%%%%%%%%%%%%%%%%%%%%%%%%%%%%%%%%%%%%%%%%%%%%%%%%%%%%%%%%%%%%

Having gone through important prerequisites, we now turn to
the analysis of the main data. In order  to determine 
a set of relatively small exponents, $\Delta_q$, to an accuracy considerably 
better than 1\%, we have developed the following procedure.
In each panel of Fig.~\ref{f2} we plot for fixed $q$ a family of 
curves labelled by a parameter $\delta$,  
\begin{equation}
F_q(N) = P_q N^{q-1 + \delta(1-q)q}.
\end{equation}
For the particular family member, 
for which $F_q(N)$ becomes independent of $N$  
in the limit of large $N$, we can conclude that 
$\delta{=}\gamma_q/d$. From such a procedure we extract the 
function $\gamma_q$ without having to resort to any 
(multi-parameter) fitting procedure. 
Similarly, by studying yet another family of curves, 
\begin{equation}
\tilde{F}_{q}(N) = L_q +\ln N [1+(1-2q)\tilde{\delta}]\:,
\label{e13}
\end{equation}
%By finding $\delta$ from the condition of saturation of $F'_q(N)$ 
we have direct access also to the function $\tilde{\gamma}_q$, 
see Eq. (\ref{e9}), without the need for numerical differentiation.

\begin{figure}[htb]
%\begin{center}
%\includegraphics[scale = 1.0]{ps/Delta.pq.eps}
%\includegraphics[width=1.0\linewidth]{ps/Delta.pq.eps}
\includegraphics[width=1.0\linewidth]{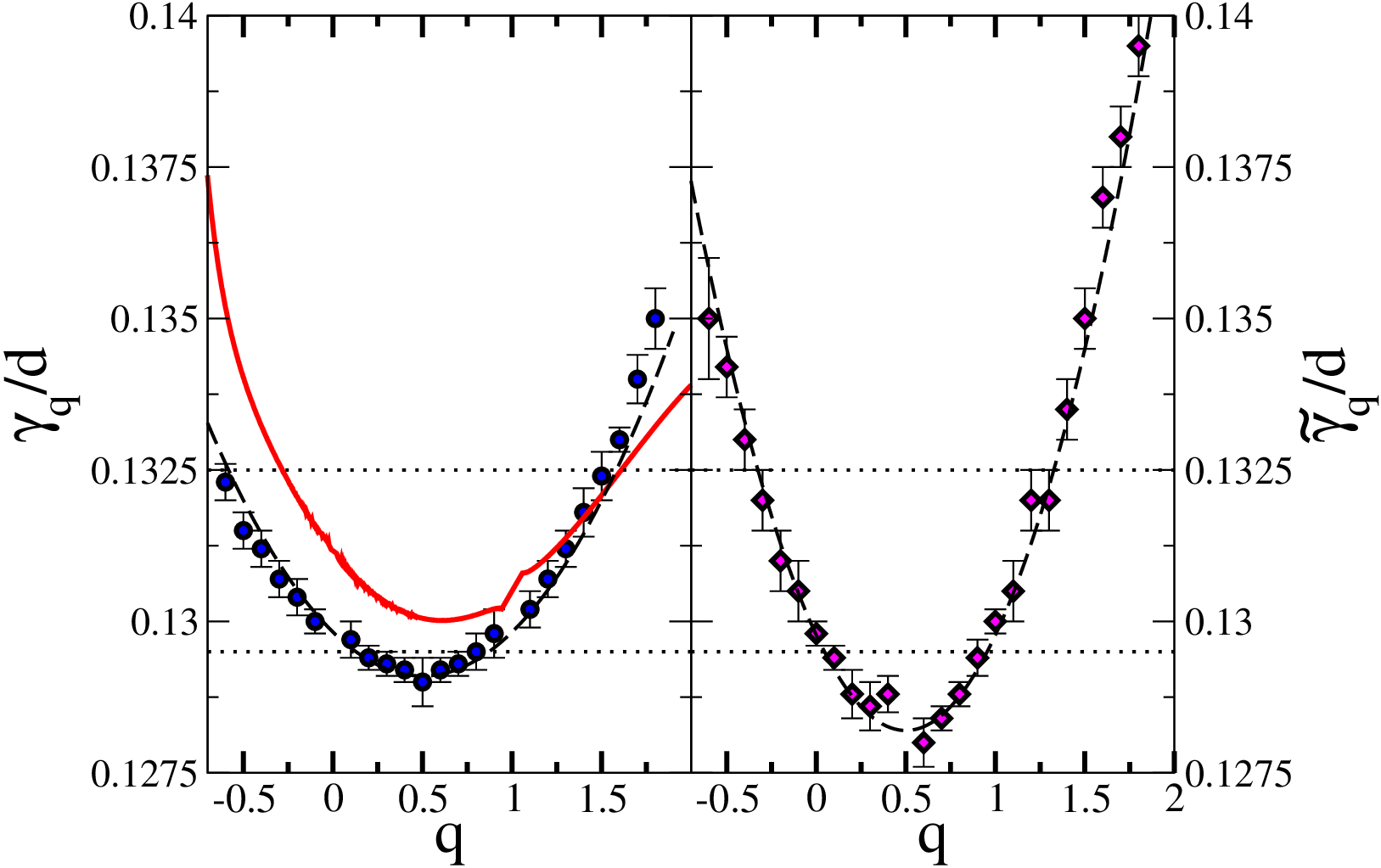}
\caption{The exponents $\gamma_q{=}\Delta_q/q(1-q)$ ($\circ$)
and $\tilde{\gamma}_q{=}\Delta'_q/(1-2q)$ ($\diamond$) 
obtained from Fig.~\ref{f2} and analogous analysis for other values of $q$.
The curvature in  $\gamma_q$ and  $\tilde{\gamma}_q$ implies that the
multifractal spectrum $\Delta_q$ is not parabolic.  
Also shown are results of the earlier work \cite{evers01}
%for the exponents obtained from average %and typical ($\circ$) 
%participation ratios 
(solid line). 
Dotted horizontal lines indicate earlier error bars in the regime 
$0\leq q \leq 1$. Dashed lines represent parabolic fits with 
$b_0{=}0.1291\pm 0.0002$, $b_1{=}0.0029\pm0.0003$
(left) and $a_0{=}0.1282\pm 0.0001$, $a_1{=}0.0063\pm0.0005$ (right).
Combination of these data yields a rough estimate of the quartic term,
$b_2 = -0.001 \pm 0.001$.}
%\end{center}
\label{f3}
\end{figure}

The functions $\gamma_q$ and $\tilde\gamma_q$ representing the main result of
this paper are  displayed  in Fig.~\ref{f3}. 
Also shown is $\Delta_q/q(1-q)$ as derived from 
the earlier evaluation of the exponents \cite{evers01} (the size of 
corresponding error bars is indicated by dotted lines). 
Figure \ref{f3} clearly shows that the curvature in
$\gamma_q$, which apparently has already left its trace 
in the earlier data, now fully reveals itself thanks to the 
reduced error bars. Even more pronounced is the 
resulting structure in the derivative $\tilde\gamma_q$. 
It is reassuring to notice that the new data confirm 
our previous finding for $b_0$ but provide a much better accuracy:  
$b_0{=}0.1291\pm 0.0002$. Our new result for the curvature of $\gamma_q$ is
$b_1{=}0.0029\pm0.0003$, which is clearly non-vanishing. These results are in
full agreement with those obtained by a fit to $\tilde{\gamma}_q$ (see the
caption to  Fig.~\ref{f3}). We thus conclude that, although the curvature of
$\gamma_q$ is numerically rather small ($b_1$ is approximately 50 times
smaller than $b_0$), it is non-zero: the multifractality spectrum $\Delta_q$
of the quantum Hall transition is not parabolic.

%%%%%%%%%%%%%%%%%%%%%%%%%%%%%%%%%%%%%%%%%%%%%%%%%%%%%%%%%%%%%%%%%%%%%%%
%%%%%%%%%%%%%%%%%%%%%%%%%%%%%%%%%%%%%%%%%%%%%%%%%%%%%%%%%%%%%%%%

\begin{figure}[t]
\vspace*{1cm}
\includegraphics[width=1.0\linewidth]{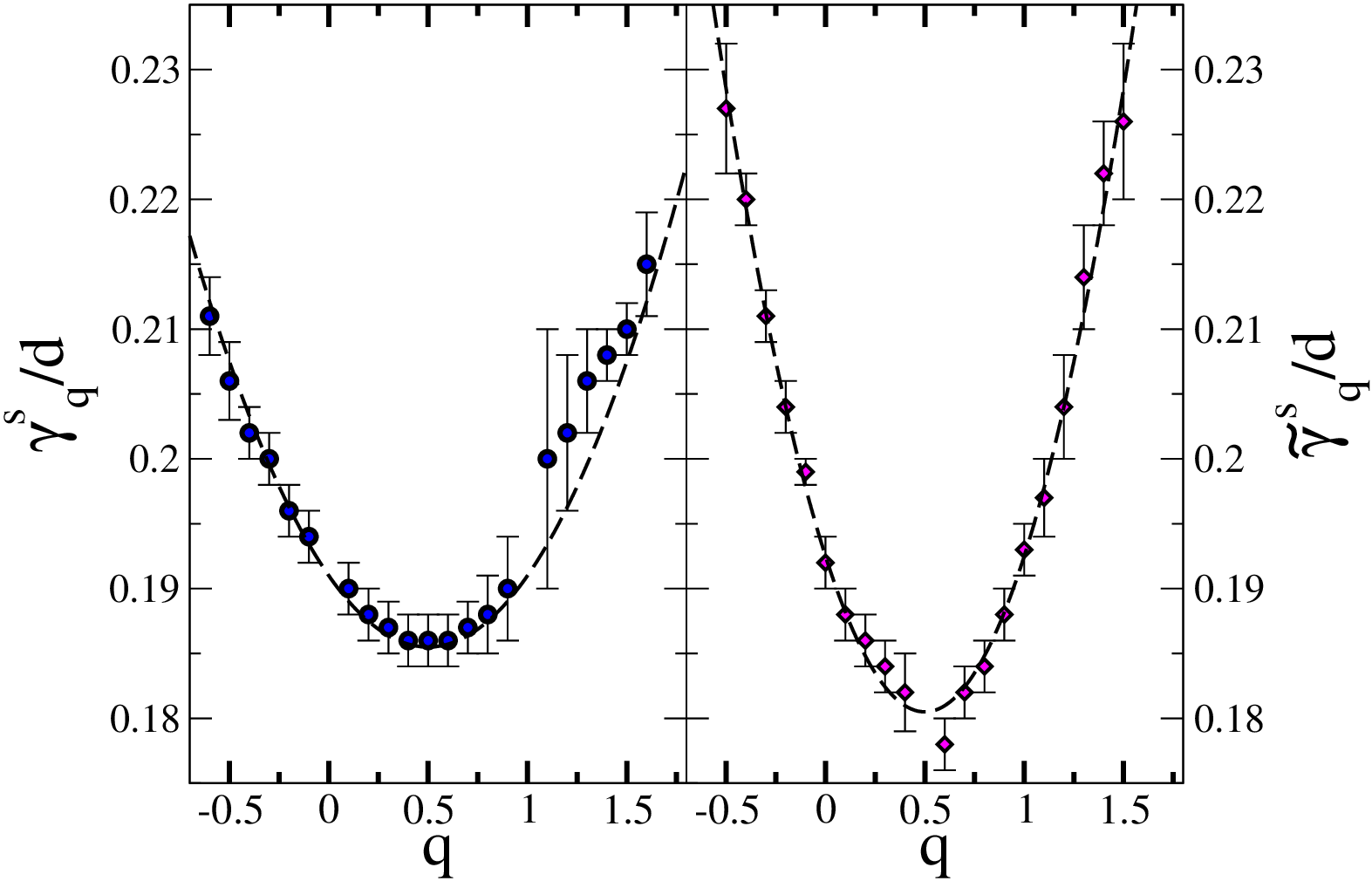}
\caption{The surface exponents $\gamma_{q}^{\rm s}{=}\Delta_{q}^{\rm
    s}/q(1-q)$. Data are presented in the form analogous to the bulk 
plot, Fig.~\ref{f3}; the curvature is even more pronounced for the surface
exponents.  
%We interpret the large error bars near $q{=}1$ as 
%a result of the denominator of $\gamma_q{=}\Delta_q/q(1-q)$
%being close do zero. 
Dashed lines indicate  parabolic fits with 
$b_0^s{=}0.1855\pm 0.0005$, $b_1^s{=}0.022\pm0.002$
(left) and $a_0^s{=}0.1805\pm 0.001$, $a_1^s{=}0.048\pm0.003$ (right).
The resulting estimate for $b_2^s$ is $b_2^s=-0.008\pm 0.01$.}
\label{f4}
\end{figure}

\paragraph*{Surface exponents:}
So far a network model with torus geometry (i.e. without boundary) 
has been considered. Recently, it has been shown \cite{subramaniam06}
that wavefunction
fluctuations near surfaces exhibit their own multifractal spectrum
with exponents $\Delta_q^{\rm s}$, defined in full analogy to 
Eq.~(\ref{e1}) via 
\begin{equation}
\label{e14}
    \langle |\psi|^{2q} \rangle_{\rm s}/\langle |\psi|^2 \rangle_{\rm s}^q \sim
    L^{-\Delta^{\rm s}_q}\:, 
\end{equation}
with the average $\langle \ldots \rangle_{\rm s}$ performed over the
boundary sites only. In general, the surface exponents $\Delta_{q}^{\rm s}$ 
are not related to their bulk counterparts in any simple manner.
We parametrize the surface spectrum in analogy with the
bulk case, Eqs.~(\ref{e2}), (\ref{e3a}), (\ref{e9}), labelling the
corresponding parameters by a superscript ``s''. The results for $\gamma_q^s$
and  $\tilde{\gamma}_q^s$ are shown in Fig.~\ref{f4}. It is seen that the
non-parabolicity of the multifractality spectrum (difference of $\gamma_q^s$
and  $\tilde{\gamma}_q^s$ from a constant) is present at the boundary as well 
and is in fact considerably more pronounced that in the bulk. The ratio
$R_q{=}\gamma_{q}^{\rm s}/\gamma_{q}$ has a clear $q$-dependence, with a
minimum at the symmetry point $q=1/2$, where it takes the value
$R_{1/2}{=}1.434\pm0.005$. It is also worth noticing that $R_q$ is appreciably
smaller than 2, a value naturally expected for critical theories expressed
in terms of a free bosonic field.

%%%%%%%%%%%%%%%%%%%%%%%%%%%%%%%%%%%%%%%%%%%%%%%%%%%%%%%%%%%%%%%%%%
%%%%%%%%%%%%%%%%%%%%%%%%%%%%%%%%%%%%%%%%%%%%%%%%%%%%%%%%%%%%%%%%%%%%%

\paragraph*{Conclusions:}
To summarize, we have studied numerically the wave function statistics at the
quantum Hall critical point. We have verified that the reciprocity relation
(\ref{e3}) holds and have used it to control systematic errors related to the
finite-size effects. In combination with a very large size of the statistical
ensemble, this has allowed us to reach unprecedented accuracy in determination
of the multifractality spectrum $\Delta_q$, 
with the error bars reduced by almost an
order of magnitude compared to the earlier work. 
The result shown in Fig.~\ref{f3} 
reads $\Delta_q = 2q(1-q)[b_0 + b_1(q-1/2)^2 + b_2(q-1/2)^4 + \ldots]$,
with  $b_0 = 0.1291\pm 0.0002$, $b_1 = 0.0029\pm 0.0003$, and $b_2=-0.001\pm
0.001$. The obtained spectrum shows clear non-parabolicity, $b_1\ne 0$, 
thus excluding a broad class of theories of the
Wess-Zumino-Witten type as candidates in the conformal field theory of the
quantum Hall transition. These results are corroborated by the analysis of the
surface mutlifractality. While completing this work, we learnt about an
independent study by Obuse {\it et al.} \cite{obuse08} who focussed on the
surface multifractality spectrum and came to the same conclusions. 
 
We thank  A.~Furusaki, I.~A. Gruzberg, A.W.W.~Ludwig, H. Obuse, and
A.~R. Subramaniam for useful discussions and for sharing their data
prior to publication. We are also grateful to A.~Tsvelik and 
M.R.~Zirnbauer for instructive discussions. This work was supported by
the Center for Functional Nanostructures of the DFG.

%\end{references}

\begin{thebibliography}{99}
\bibitem{note1} More accurately, the parabolicity may only hold for 
$q<q_c$; its termination at $q=q_c\equiv ... $ is related to wave function
normalization, see the review  \cite{evers07} for detail.
\end{thebibliography}
\end{document}